\documentclass[10pt]{article}

\usepackage{booktabs}
\usepackage{subfig}
\usepackage{tabularx}

\usepackage{graphicx}
\usepackage{epstopdf}
\usepackage{layouts}
\usepackage[bottom]{footmisc}
\usepackage{authblk}

\usepackage{listings}
\usepackage{color}

\definecolor{dkgreen}{rgb}{0,0.6,0}
\definecolor{gray}{rgb}{0.5,0.5,0.5}
\definecolor{mauve}{rgb}{0.58,0,0.82}

\begin{document}

\title{Synthetic Data Generation using Benerator Tool}
\author[1]{Vanessa Ayala-Rivera\thanks{vanessa.ayala-rivera@ucdconnect.ie}}
\author[2]{Patrick McDonagh\thanks{patrick.mcdonagh@dcu.ie}}
\author[1]{Thomas Cerqueus\thanks{thomas.cerqueus@ucd.ie}}
\author[1]{Liam Murphy\thanks{liam.murphy@ucd.ie}}
\affil[1]{Lero@UCD, Performance Engineering Laboratory, 
School of Computer Science and Informatics, University College Dublin}
\affil[2]{Lero@DCU, Performance Engineering Laboratory, 
School of Electronic Engineering, Dublin City University}
\renewcommand\Authands{ and }
\maketitle
\centerline{University College Dublin, Technical Report UCD-CSI-2013-03}

\begin{abstract}
Datasets of different characteristics are needed by the research community for
experimental purposes. However, real data may be difficult to obtain due to
privacy concerns.
Moreover, real data may not meet specific characteristics which are needed to
verify new approaches under certain conditions. Given these limitations,
the use of synthetic data is a viable alternative to complement the real
data.
In this report, we describe the process followed to generate synthetic
data using Benerator, a publicly available tool. The results show that the synthetic
data preserves a high level of accuracy compared to the original data.
The generated datasets correspond to microdata containing records with social,
economic and demographic data which mimics the distribution of aggregated
statistics from the 2011 Irish Census data. 
\end{abstract}

\section{Introduction}
The creation of synthetic data is an approach widely used by the research
community in a variety of domains: privacy protection~\cite{Machanavajjhala2008,
Ghinita2008}, healthcare \cite{Cooley2010}, pattern recognition
\cite{Varga2003}, data mining \cite{Liu2011}, etc.
Such data is often generated to meet specific characteristics that are not
found in the real data. By generating synthetic datasets, researchers can
have more flexibility on the manipulation of the data and are able to test a
wider set of conditions and scenarios in their applications. 
Moreover, synthetic data is also used as a substitute for real data, as it
is often difficult to obtain due to privacy concerns (i.e., to
protect the privacy of the individuals represented in the real data).

It is often the case that researchers adopt the practice of generating
synthetic data to test their proposed algorithms against more heterogeneous
datasets. In particular, some of the reasons why we believe the generation of
synthetic data is a useful approach for the research community are the
following:

\begin{itemize}
	\item Generating synthetic data allows to control the data distributions used
	for testing. One can study the behavior of the algorithms under different
	conditions: identify scenarios where the data distribution favors the
	performance of the algorithm or the scenarios where the algorithm performs the
	worst. 
	\item Synthetic data can help to allow a fair performance comparison
	among the algorithms. For example, for evaluating the scalability of the
	algorithms. When the same dataset is used for testing, but only scaling up its size while preserving
	the same data distribution, the measures obtained in terms of efficiency (e.g.,
	memory consumption, running time) give a more precise idea about the
	causes for the increase of the computational resources. Especially for
	algorithms that could be affected by some aspects of the datasets like the cardinality of
	the attributes.
	\item Generating synthetic data allows to create records which have the
	finest level of granularity in each attribute. In contrary, publicly available
	real datasets have often undergone anonymization procedures due to privacy
	constraints. Therefore, the values for some of the attributes are already
	grouped in less specific values. For example, when some values are sparse in
	the dataset, they are all placed together in a group to protect the privacy of
	the individuals that fall in that minority population.
\end{itemize}

In order to benefit from this approach, it is important to have practical tools
that can be customized and easily extended according to different needs.

In this work we present a practical approach to generate synthetic census-based
data. We describe the methodology followed to generate this type of datasets and
how to simplify its creation using Benerator \cite{beneratorTool}, a publicly
available tool.

\section{Synthetic Data Generation}

In this section we describe the process followed to generate microdata
\cite{Willenborg2001}, where the records correspond to the information about an
individual. The process can be applied to microdata of different domains.
However, in our work, we have applied it to create personal data records. Each
record contains information about social, economic and demographic attributes of
a person. This data is based on attributes collected from the Irish Census
2011\footnote{http://www.cso.ie/en/databases/}.

Given that access to microdata is commonly restricted, the census publishes only
the aggregated statistics for some of the attributes. Often, these attributes
are correlated with each other and the statistics for these relationships are
disseminated. One example of the published relationships are: marital status by
age group; and highest level of education completed by socio-economic group.
In our analysis of census data, we capture the frequency distributions of
multiple attributes and use them as probability weights in the data generation
process. By using this approach, we preserve the density of the
population corresponding to the selected demographic attributes.

We captured the statistics from people corresponding to the adult
population only (i.e., ages between 17 and 84 years). This selection results in
a population count of 3,550,246 people. This is the total number of records we
considered as the original Irish Census data in our work to verify the
accuracy in which the synthetic data is generated.

\subsection{Overview of the Generation Process}
There are different frameworks to generate synthetic data
\cite{dbmonsterTool, generatedataTool, jailerTool}.
In our case we used the Java open-source tool Benerator \cite{beneratorTool}, as
it offers high capabilities of extensibility and customization.
We implemented the logic to use the census data as the domain values and the
aggregated statistics as weights for the generation process.

\begin{figure}
\centering
\includegraphics[height=3.5in, width=4in]{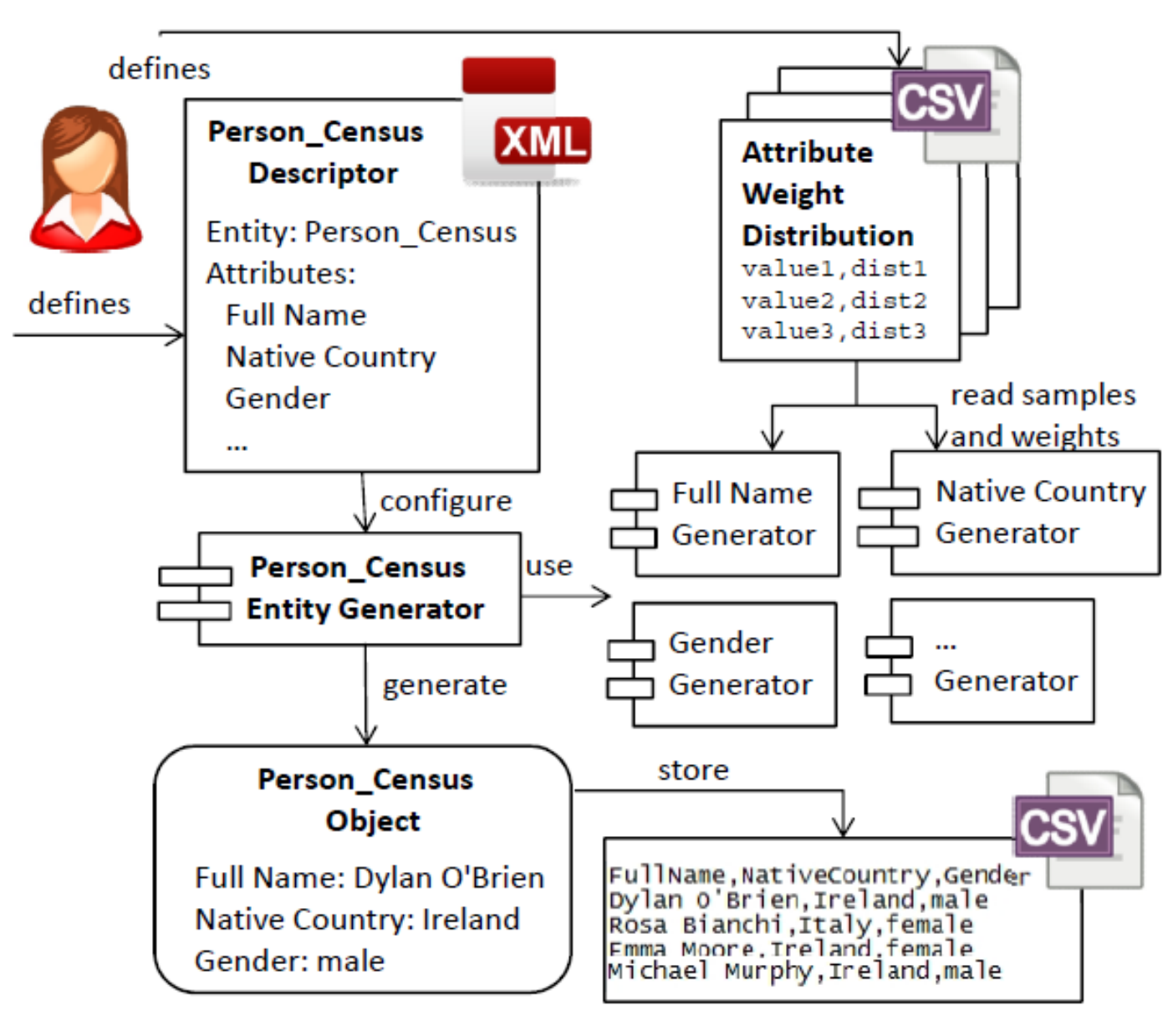}
\caption{Overview of the synthetic data generation process.}
\label{fig:syntheticDataGenProcess}
\end{figure}

Figure~\ref{fig:syntheticDataGenProcess} shows the steps and components involved
in the generation of the synthetic data. Firstly, we configure a descriptor
file, which is stored in XML format. In this file we indicate
the number of records to be generated, the type of entity to be created (e.g., a
census-based person) and its attributes (e.g., \textit{age,
gender, marital status}); and the type of output for the dataset (e.g., plain
text file, databases).
In a separate file, which is stored in comma-separated value
format (CSV), we also configure the universe of values that each
attribute can take and the distribution they will follow in the dataset.
Whereas the records can be randomly generated, in our case, we wanted to mimic
the distribution from the aggregated statistics from the Irish Census.

Once the metadata has been set up, the generator object for the entity
(i.e., \textit{entity generator}) will start the creation of the synthetic
records. The \textit{Entity generator} acts as the controller to run the
\textit{sub generators} corresponding to each of the demographic attributes:
Once the entity object is created, it will be made available to the other
\textit{sub generators} so they can use it in their generation process.
The \textit{entity generator} has also the logic to respect a set of constraints
between some of the attributes (one of the main characteristics of realistic
datasets). This logic ensures to assign realistic values for the attributes.
For example, when generating \textit{marital status}, it is important to
consider that most of the population between 15 and 19 years are single thus
the \textit{generator} should not assign a widowed status, which is less likely.

The outcome of the synthetic generation process is a set of entities (e.g,
census-based person records) which are stored in CSV files. We chose to use CSV
format because it is widely-known format, easy to manipulate (i.e.,
export/import) and portable among different types of applications (e.g., supported by most of
frameworks and tools).

\subsection{Implementation Details}

\begin{figure}[!t]
\centering
\includegraphics[height=5in, width=5in]{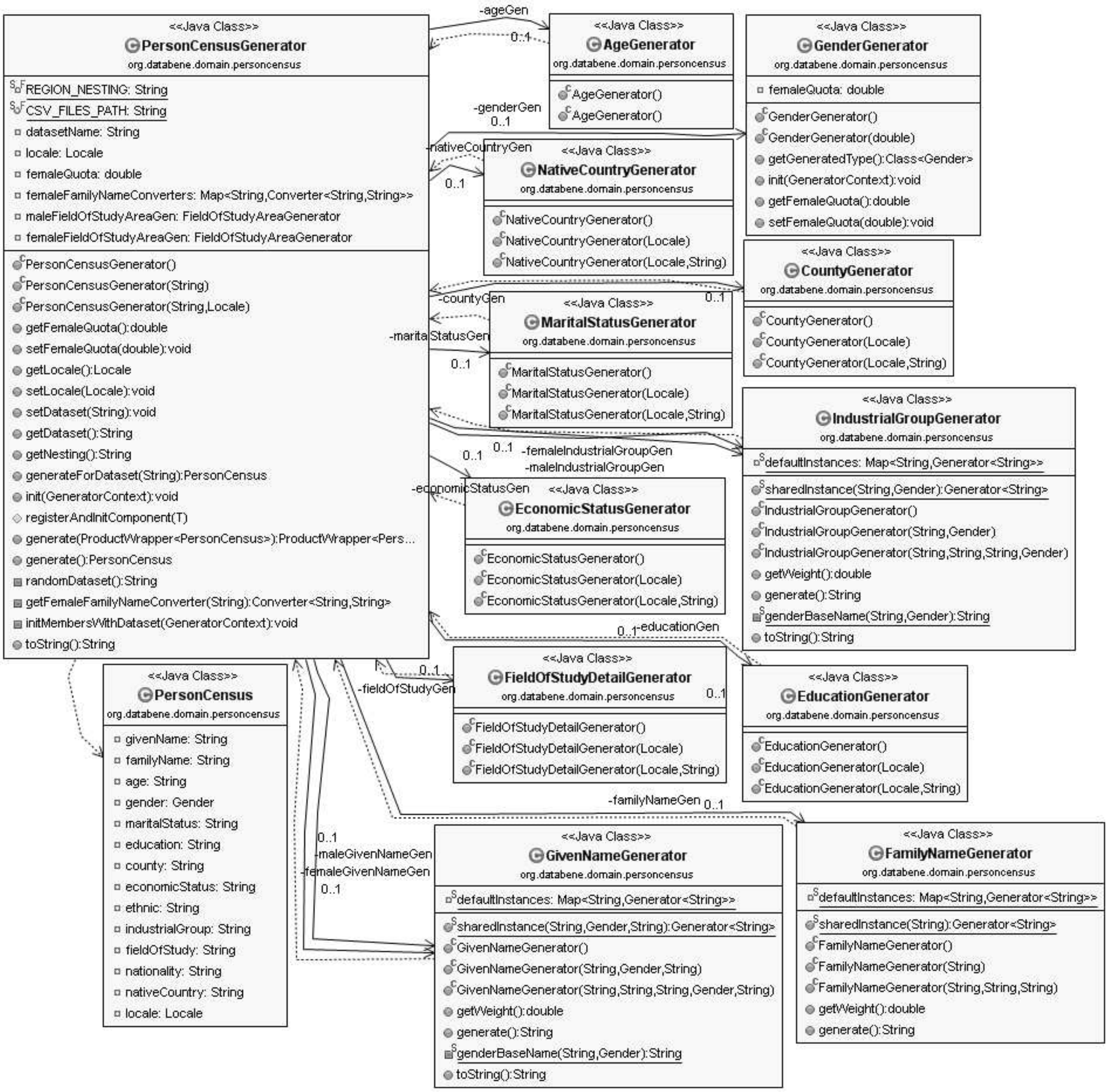}
\caption{Class diagram for the generation of person census records.}
\label{fig:classDiagram}
\end{figure}

We customized Benerator to generate personal records according to the
Irish Census data. Figure~\ref{fig:classDiagram} shows the classes
involved in our customization.
In order to respect the multi-attribute constraints, we followed a specific
order in the generation of the data that is part of the \textit{personcensus}
entity. This logic is implemented in the \textit{PersonCensusGenerator} class.
The first attributes to be generated are the independent ones:
\textit{gender}, \textit{nationality} and \textit{age}. Once the values for
those attributes have been generated, the dependent attributes can use them to
generate their own data.
For example, \textit{full name} is dependent on \textit{gender} and
\textit{nationality} to be customized accordingly. \textit{Industrial group}
and \textit{field of study} are dependent on \textit{age} and \textit{gender}. The
rest of the attributes (\textit{education}, \textit{marital status}, etc.) are
generated using \textit{age} groups as base.

\newpage
To better illustrate the synthetic generation process, we provide an example
of the configuration files we defined in our work.
\newline

\textbf{Descriptor file:} We configured the location as
\textit{IE} to indicate Benerator to use the set of files regionalized for
\textit{Ireland}. The type of entity to generate is setup as
\textit{PersonCensus} which is the name for the generator class. The
number of entities to be created is set to 30,000. The generated entity is
kept in a variable with name \textit{person}. All the attributes that are
part of this entity are listed with their corresponding sub generator. Finally,
the output is directed to a CSV file. An example of a descriptor file is shown
in \texttt{PersonCensusDescriptor.xml}.
\newline
\newline

\begin{lstlisting}
// PersonCensusDescriptor.xml
<?xml version="1.0" encoding="iso-8859-1"?>
<setup defaultDataset="IE">
    <import domains = "personcensus" />
    <generate type="PersonCensus" count="30000" >
        <variable name="person" generator="PersonCensusGenerator" dataset="IE" locale="IE"/> 
        <attribute name="gender" script="person.gender" />	
        <attribute name="age" script="person.age" />					
        <attribute name="maritalStatus" script="person.maritalStatus" />		        
        <attribute name="economicStatus" script="person.economicStatus"/>       
        <attribute name="NativeCountry" script="person.nativeCountry" />		
        <attribute name="FullName" script="person.givenName + ' ' +
                	person.familyName" /> 
        <consumer class="org.databene.platform.csv.CSVEntityExporter">
            <property name="uri" value="./output/irishcensus30m.csv"/>            
            <property name="columns" value="FullName, NativeCountry, 
            		Gender,	Age, MaritalStatus, EconomicStatus"/> 
        </consumer>
    </generate> 
</setup>
\end{lstlisting}

\textbf{Configuration CSV files:} We defined the universe of values for each
attribute and the weights to be used by the distribution function in Benerator.
For the independent attributes, like \textit{nationality}, a single file is
configured (as shown in file \texttt{Nationality.csv}). In contrary, for
dependent attributes, like \textit{marital status} which depends on \textit{age}, one file is configured
for each \textit{age group} (as shown in files
\texttt{MaritalStatusQty15-19.csv} and \texttt{MaritalStatusQty80-84.csv}).
\begin{lstlisting}
//Nationality.csv
Irish,969087
Austrian,554
Belgian,932
Bulgarian,865
Cypriot,53
Czech,1921
...
Australian,2190
New Zealander,914

//MaritalStatusQty15-19.csv
Single,282106
Married (first marriage),836
Re-married (following widowhood),4
Re-married (following dissolution of previous marriage),3
Separated (including deserted),55
Divorced,8
Widowed,7

//MaritalStatusQty80-84.csv
Single,11898
Married (first marriage),25133
Re-married (following widowhood),565
Re-married (following dissolution of previous marriage),209
Separated (including deserted),699
Divorced,308
Widowed,31301
\end{lstlisting}

\begin{figure}[!h]
\centering
\includegraphics{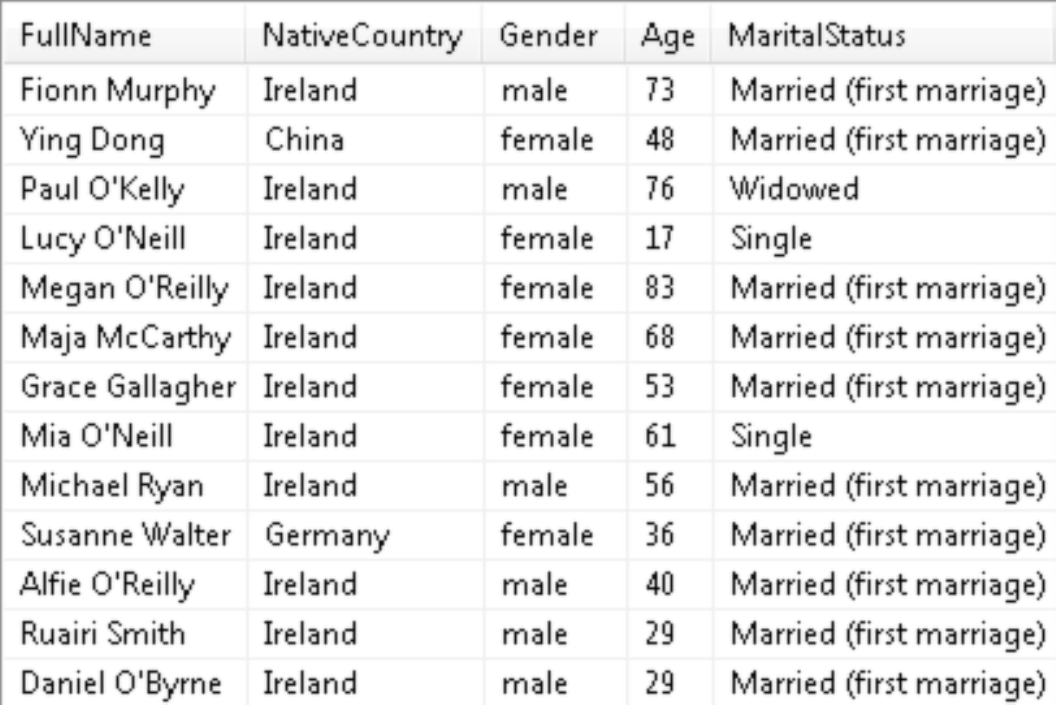}
\caption{Example of generated data for person census domain.}
\label{fig:outputCSV}
\end{figure}

\textbf{Generated Dataset (CSV file):} In Figure~\ref{fig:outputCSV}, we show an
example of a dataset generated based on the Irish Census.
\section{Results}
\label{sec:results}

In this section we describe the generated datasets: detailing the attributes
which are part of the datasets and the domain values for each attribute.
Moreover, we show the comparison between the data distributions from the
Irish Census data and the distribution from a dataset that was
synthetically generated based on the aggregate statistics of the census data.
The original census dataset was composed of 3,550,246 records and we scaled down
the dataset population to different sizes. The results shown in this report are
those obtained for a dataset with 100k records.

\subsection{Comparison of Real and Synthetic Datasets}
\subsubsection{Independent Attributes}

\begin{figure}[!h]
\centering
\includegraphics[height=2in, width=5in]{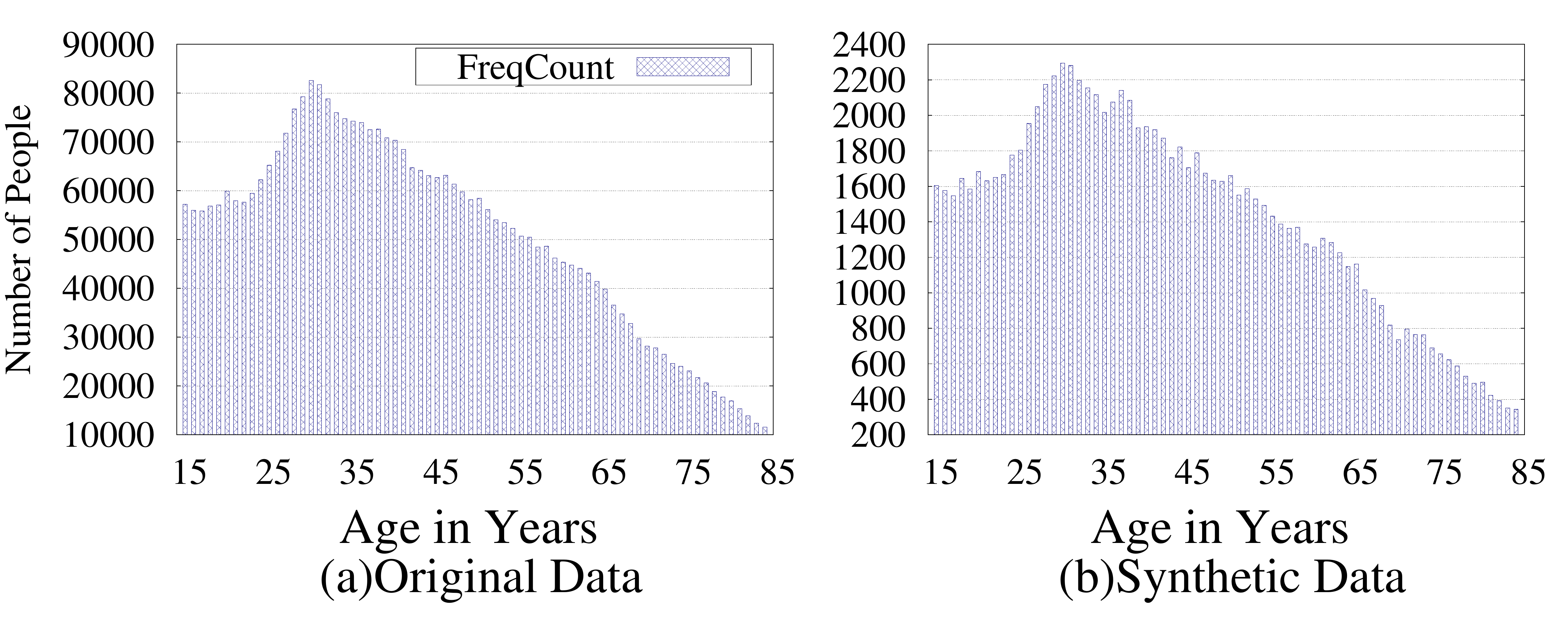}
\caption{Comparison of data distribution between original and synthetic
dataset for the \textit{age} attribute.}
\label{fig:ageComparison}
\end{figure}

Figure~\ref{fig:ageComparison} shows the comparison between the
distribution from the Irish Census data (a) and the synthetic dataset
(b) for the \textit{age} attribute. It can be seen that the
data distribution between the synthetic and the original data are
similar, showing that the generation process preserves a good level of accuracy
from the original distribution.
\subsubsection{Dependent Attributes}

\begin{figure}[!h]
\centering
\includegraphics[height=4.2in, width=5in]{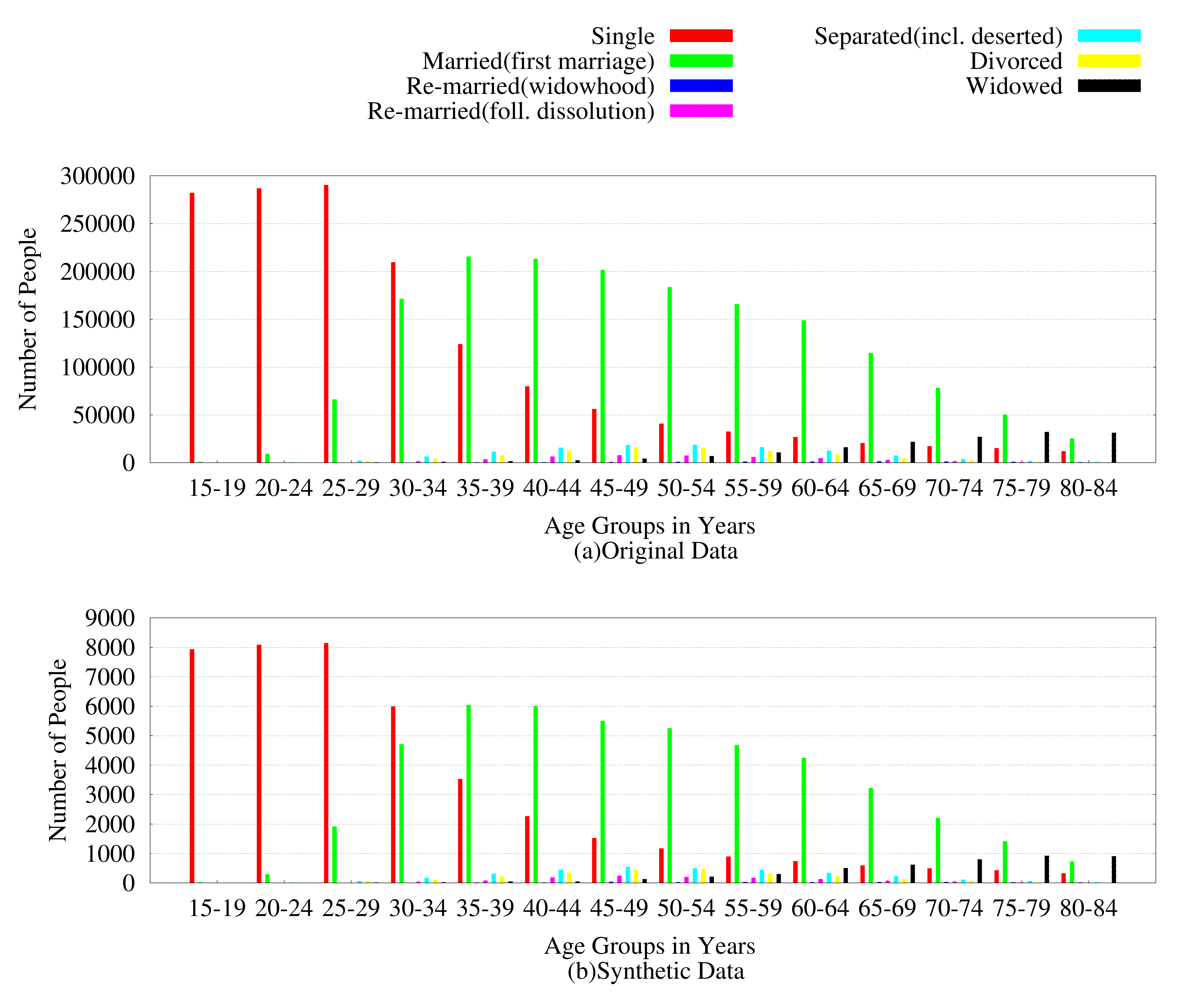}
\caption{Comparison of data distribution between original and synthetic dataset for
constraint attribute \textit{marital status per age groups}.}
\label{fig:maritalComparison}
\end{figure}

Figure~\ref{fig:maritalComparison} shows the comparison between the
distribution from the Irish Census data (a) and the synthetic dataset
(b) for the \textit{marital status} attribute, which is constrained by
\textit{age group}. It can be seen that the generated synthetic data
preserves a high level of accuracy compared to the original distribution
among the \textit{age} groups.
\section{Synthetic Datasets Description}
\label{sec:datasetdesc}

The generated datasets are formed by the following attributes: 

\begin{enumerate}
	\item {\sl Full name:} These are randomly generated names that result from a
	concatenation of a list of given names and family names. Full names are
	generated according to the nationality and age attributes to match the most
	commonly used names in those countries.
	\item {\sl Age:} Values between 17 and 84 in order to only consider the adult
	population.
	\item {\sl Gender:} female, male. 
	\item {\sl County:} The list of all Irish counties which are: Carlow, Dublin 
	City, D�n Laoghaire-Rathdown, Fingal, South Dublin, Kildare, Kilkenny, Laois, Longford, Louth, Meath, Offaly, Westmeath,
	Wexford, Wicklow, Clare, Cork City, Cork County, Kerry, Limerick City, Limerick
	County, North Tipperary, South Tipperary, Waterford City, Waterford County,
	Galway City, Galway County, Leitrim, Mayo, Roscommon, Sligo, Cavan, Donegal,
	Monaghan. 
	\item {\sl Marital Status:} Single, Married (first marriage), Re-married
	(following widowhood), Re-married (following dissolution of previous marriage),
	Separated (including deserted), Divorced, Widowed.
	\item {\sl Native Country:} Ireland, Austria, Belgium, Bulgaria, Cyprus, Czech
	Republic, Denmark, Estonia, Finland, France, Germany, Greece, Hungary, Italy,
	Latvia, Lithuania, Luxembourg, Malta, Netherlands, Poland, Portugal, Romania,
	Slovakia, Slovenia, Spain, Sweden, Russian Federation, Ukraine, Niger, South
	Africa, Mauritius, India, Philippines, China, Pakistan, Malaysia, United States
	of America, Brazil, Canada, Australia, New Zealand.
	\item {\sl Economic Status:} Employer or own account worker, Employee,
	Assisting relative, Unemployed looking for first regular job, Unemployed having
	lost or given up previous job, Student or pupil, Looking after home/family,
	Retired, Unable to work due to permanent sickness or disability, Other economic
	status.
	\item {\sl Industrial Group:} Agriculture, forestry and fishing (A), Mining and
	quarrying (B), Manufacturing (C), Electricity, gas, steam and air conditioning
	supply (D), Water supply; sewerage, waste management and remediation activities
	(E), Construction (F), Wholesale and retail trade; repair of motor vehicles and
	motorcycles (G), Transportation and storage (H), Accommodation and food service
	activities (I), Information and communication (J), Financial and insurance
	activities (K), Real estate activities (L), Professional, scientific and
	technical activities (M), Administrative and support service activities (N),
	Public administration and defence; compulsory social security (O), Education
	(P), Human health and social work activities (Q), Arts, entertainment and
	recreation (R), Other service activities (S), Activities of households as
	employers producing activities of households for own use (T), Activities of
	extraterritorial organisations and bodies (U).
	\item {\sl Education:} No formal education, Primary, Lower secondary, Upper
	secondary, Technical/vocational, Advanced certificate/completed apprenticeship,
	Higher certificate, Ordinary bachelor degree/professional qualification or
	both, Honours bachelor degree/professional qualification or both, Postgraduate
	diploma or degree, Doctorate (Ph.D).
	\item {\sl Field of Study:} Education and teacher training, Music and
	performing arts, Audio-visual techniques and media production, Design, Other
	arts, Foreign languages, Mother tongue, History and archaeology, Other
	humanities, Psychology, Economics, Business and administration (broad
	programmes), Marketing and advertising, Accounting and taxation, Management and
	administration, Secretarial and office work, Law, Other social sciences,
	business and law subjects, Biology and biochemistry, Physical sciences
	(physics, chemistry, earth science), Computer science, Computer use, Other
	science, mathematics and computing, Engineering and engineering trades (broad
	programmes), Mechanics and metalwork, Electricity and energy, Motor vehicles,
	ships and aircraft, Architecture and town planning, Building and civil
	engineering, Other engineering, manufacturing and construction, Crop and
	livestock production, Other agriculture and veterinary, Medicine, Nursing and
	caring, Child care and youth services, Social work and counselling, Other
	health and welfare, Hotel, restaurant and catering, Hair and beauty services,
	Other personal services, Air transportation, Ground transportation, Sea
	transportation, Other transportation services, Public security services,
	Industrial security services, Other security services, Other subjects.
\end{enumerate}

\section{Conclusions}
\label{sec:conclusions}

In this report we described the approach we used to generate synthetic
microdata. We described how we extended the public available tool, called
Benerator, to add a new domain for data generation using census-based personal
records. We showed how capturing the frequency distributions from a census
dataset and using them as probability weights in the generation process, we were
able to mimic the data distribution for that census. We also presented the
comparison of the distributions between the original and the synthetic data,
showing that synthetic dataset preserves a high level of accuracy compared
to the original distribution.
Finally, we provided a detailed description of the generated datasets about the
domain values used for each attribute.
 
\section*{Acknowledgements}
This work was supported, in part, by Science Foundation Ireland grant
10/CE/I1855 to Lero - the Irish Software Engineering Research Centre
(www.lero.ie). This work has also received support from Science Foundation
Ireland (SFI) via grant 08/SRC/I1403 FAME SRC (Federated, Autonomic Management
of End-to-End Communications Services - Scientific Research Cluster).

\bibliographystyle{abbrv}
\bibliography{techreportbib}  

\end{document}